\documentclass[journal = ancac3, manuscript = article, layout = twocolumn, 12p]{achemso} 

\usepackage{siunitx}
\usepackage{setspace}
\author{Alberto Nardi}
\affiliation[Research Laboratory of Electronics]
{Research Laboratory of Electronics, Massachusetts Institute of Technology, 50 Vassar Street, Cambridge, MA 02139, USA}
\alsoaffiliation[Politecnico di Torino]
{Department of Electronics and Telecommunications, Politecnico di Torino, Corso Duca degli Abruzzi 24, Torino 10129, Italy}
\author{Marco Turchetti}
\affiliation[Research Laboratory of Electronics]
{Research Laboratory of Electronics, Massachusetts Institute of Technology, 50 Vassar Street, Cambridge, MA 02139, USA}
\author{Wesley A. Britton}
\affiliation[Materials Science]
{Division of Materials Science \& Engineering, Boston University, 15 Saint Mary’s Street, Brookline, MA 02446, USA}
\author{Yuyao Chen}
\affiliation[EECS]
{Department of Electrical \& Computer Engineering and Photonics Center, Boston University, 8 Saint Mary’s Street, Boston, MA 02215, USA}
\author{Yujia Yang}
\affiliation[Research Laboratory of Electronics]
{Research Laboratory of Electronics, Massachusetts Institute of Technology, 50 Vassar Street, Cambridge, MA 02139, USA}
\author{Luca Dal Negro}
\affiliation[Materials Science]
{Division of Materials Science \& Engineering, Boston University, 15 Saint Mary’s Street, Brookline, MA 02446, USA}
\alsoaffiliation[EECS]
{Department of Electrical \& Computer Engineering and Photonics Center, Boston University, 8 Saint Mary’s Street, Boston, MA 02215, USA}
\alsoaffiliation[Physics]
{Department of Physics, Boston University, 590 Commonwealth Avenue, Boston, MA 02215, USA}
\author{Karl K. Berggren}
\affiliation[Research Laboratory of Electronics]
{Research Laboratory of Electronics, Massachusetts Institute of Technology, 50 Vassar Street, Cambridge, MA 02139, USA}
\author{Phillip D. Keathley}
\affiliation[Research Laboratory of Electronics]
{Research Laboratory of Electronics, Massachusetts Institute of Technology, 50 Vassar Street, Cambridge, MA 02139, USA}
\email{pdkeat2@mit.edu}

\title{Refractory doped titanium nitride nanoscale field emitters}


\begin{document}


\begin{abstract}
Refractory materials exhibit high damage tolerance, which is attractive for the creation of nanoscale field-emission electronics and optoelectronics applications that require operation at high peak current densities and optical intensities. Recent results have demonstrated that the optical properties of titanium nitride, a refractory and CMOS-compatible plasmonic material, can be tuned by adding silicon and oxygen dopants.  However, to fully leverage the potential of titanium (silicon oxy)nitride, a reliable and scalable fabrication process with few-nm precision is needed. In this work, we developed a fabrication process for producing engineered nanostructures with gaps between 10 and \SI{15}{\nano\metre}, aspect ratios larger than 5 with almost \ang{90} steep sidewalls. Using this process, we fabricated large-scale arrays of electrically-connected bow-tie nanoantennas with few-nm free-space gaps. We measured a typical variation of \SI{4}{\nano\metre} in the average gap size. Using applied DC voltages and optical illumination, we tested the electronic and optoelectronic response of the devices, demonstrating sub-10-\si{\volt} tunneling operation across the free-space gaps, and quantum efficiency of up to $1\times10^{-3}$ at \SI{1.2}{\micro\metre}, which is comparable to a bulk silicon photodiode at the same wavelength. Tests demonstrated that the titanium silicon oxynitride nanostructures did not significantly degrade, exhibiting less than \SI{5}{\nano\metre} of shrinking of the average gap dimensions over few-\si{\micro\metre\squared} areas after roughly 6 hours of operation. Our results will be useful for developing the next generation of robust and CMOS-compatible nanoscale devices for high-speed and low-power field-emission electronics and optoelectronics applications.  
\end{abstract}

\newpage
With vacuum gaps smaller than the electron's mean free path and sharp, few-nm features, nanoscale vacuum electronics are capable of petahertz operation and thanks to vacuum-like ballistic transport have demonstrated improved stability with respect to environmental factors such as temperature and radiation \cite{Nirantar2018,02}. Furthermore, when illuminated with optical fields, these nanoscale structures can behave as sub-wavelength antennas for harvesting light for photodetection and optoelectronic applications~\cite{Herink_photoemission,Swanwick_silicon_tip_arrays,Crozier_optical_antennas,Racz_ultrafast_photoemission,Rybka_subcycle_control}.  Noble metals are often used for enhanced optical sensitivity as their negative permittivities allow them to support localized surface plasmonic (LSP) resonances that result in significant absorption and surface field enhancements.  However, the application of devices fabricated from noble metals is often limited due to their low melting points, high surface mobility, and lack of CMOS-compatibility. Specifically, noble-metal-based nanostructures are highly sensitive to laser fluence and, depending on their specific geometry, can have energies required for photothermally induced shape transformation that are much smaller than the already small energies required for nanostructure melting \cite{Inasawa_surface_melting,Walsh_second_harmonic}. These shape transformations can dramatically reduce device performance \cite{Sivis_UV_plasmonic}. For example, in Ref. \citep{Yang_Nat_comm}, we see that laser-induced reshaping of Au bow-tie nanoantennas on glass degrades the field enhancement and the optically-induced current over time.

As a possible solution, refractory plasmonic materials such as titanium nitride (TiN) have gained attention in the last few years \cite{32,Bagheri_Nb_plasmonics,Bagheri_TiN_arrays,Gui_tiN_nanoantennas} due to their high bulk melting points and, thus, robustness under optical illumination and high peak current density operation. However, refractory materials can be challenging to process, leading to the need for new etching procedures.\cite{Yamaguchi_flame_etching,Chaika_W_etching} Thin films of TiN can be grown by magnetron sputtering (MSP) with tunable optical properties similar to Au in the visible and near-IR spectral range \cite{17,18, 30, Gioti_TiN_epitaxial}. Furthermore, TiN thin films have been shown to result in enhanced cold-field emission relative to tungsten emitters~\cite{lo_titanium_1996}, and to support hot carriers with long lifetimes \cite{31}, leading to improved performance relative to metal coatings such as Au for photodetection applications \cite{cleo}. 

Recently, it has been demonstrated that the introduction of Si- and O-dopants into TiN (TiSi$_x$O$_y$N$_z$) allows for large tunability of the optical dispersion of TiN thin films, inducing anomalous dispersion behavior at visible and near-infrared wavelengths \cite{19}. Besides enabling larger flexibility for nanostructure device engineering, this anomalous dispersion behavior can support broadband optical responses by satisfying the condition for the excitation of LSPs over multiple spectral bands. Broadband optical responses are favorable for increased detector bandwidths and for the detection of ultra-short pulses and frequency combs, both of which intrinsically contain a wide spectrum of frequency contributions.

To fully leverage the unique material properties of TiSi$_x$O$_y$N$_z$ films for nanoscale electronic and optoelectronic applications, a reliable and scalable fabrication process with few-nm precision needs to be developed. Good control over gap dimensions and sidewall steepness is needed to optimize the current emission across the gap. So far, gaps of the order of \SI{20}{\nano\metre} have been recently achieved in TiN bow-tie nanoantennas \cite{Gadalla2019TiN_excitation}. TiSi$_x$O$_y$N$_z$ films add another challenge to the fabrication process since the stoichiometry of the doped films combines the chemical and material properties of titanium nitride, titanium oxide, and silicon oxide. Therefore, a common etchant to the different materials is needed to be able to successfully pattern TiSi$_x$O$_y$N$_z$. 

\begin{figure*} [h!]
\centering
  \includegraphics[width=1\linewidth]{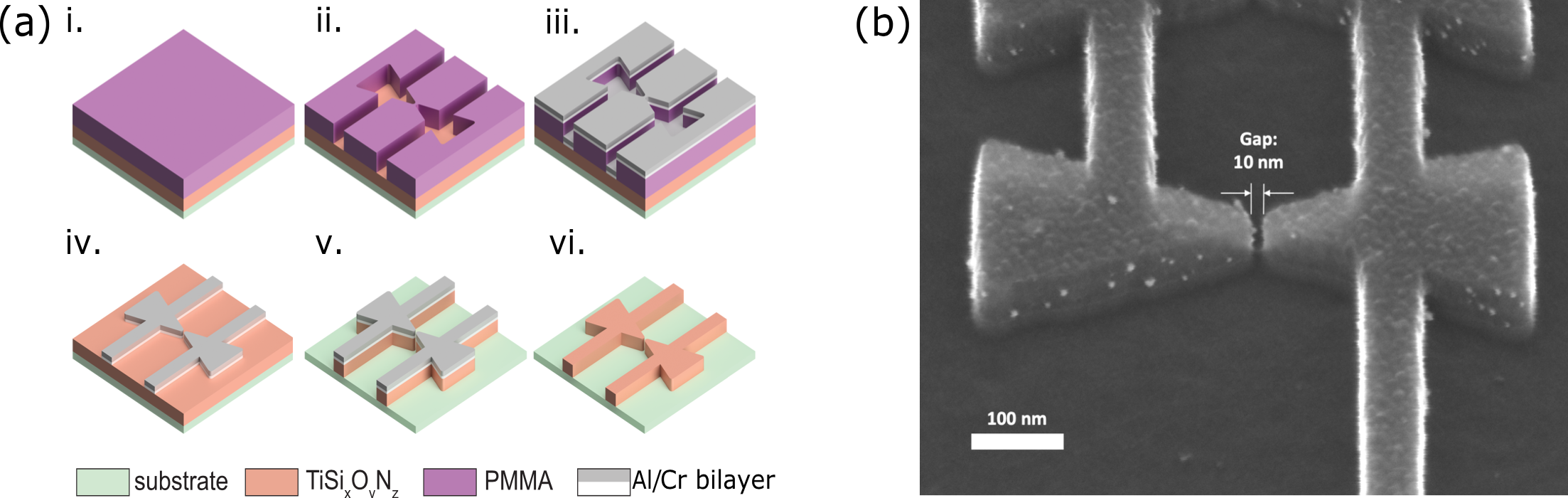}
  \caption{(a) Fabrication process for patterning TiSi$_x$O$_y$N$_z$ nanostructures. (a.i) Spin of e-beam positive resist. (a.ii) Electron beam exposure and development. (a.iii) Evaporation of a Al/Cr bilayer hard mask. (a.iv) Resist lift-off. (a.v) Reactive ion etching of TiSi$_x$O$_y$N$_z$. (a.vi) Hard mask removal by wet etching. (b) \ang{45}-tilted scanning electron micrograph of a fabricated TiSi$_x$O$_y$N$_z$ bow-tie nanoantenna.}
  \label{fig:Figure1}
\end{figure*}

In this work, we developed a fabrication process for producing engineered nanostructures having between 10- and 15-\si{\nano\metre}-wide gaps, aspect ratios (\textit{i.e.} the ratio between the film thickness and the gap dimension) of around 5 with almost \ang{90} sidewalls (see Figure \ref{fig:Figure1}(a)). Using this process, we fabricated arrays of electrically-connected bow-tie nanoantennas with few-nm free-space gaps (see Figure \ref{fig:Figure1}(b)). We measured a typical variation of \SI{4}{\nano\metre} in the average gap size. Using applied DC voltages and optical illumination, we tested the electronic and optoelectronic response of the devices, demonstrating sub-10-\si{\volt} static tunneling operation across the free-space gaps, and a bias-enhanced optoelectronic response having a measured quantum efficiency of up to $1\times10^{-3}$ at \SI{1.2}{\micro\metre}.  Post-analysis demonstrated that the TiSi$_x$O$_y$N$_z$ nanostructures experienced only minor degradation during operation, exhibiting less than \SI{5}{\nano\metre} of shrinking of the average gap dimensions over few-\si{\micro\metre\squared} areas after roughly 6 hours of operation, unlike gold nanorod and nanoantenna devices which are often damaged by intense optical radiation or the emission of large current densities \cite{16, Yang_Nat_comm}. Our work paves the way toward robust, long-lifetime vacuum nanoelectronics and nano-optoelectronics capable of low-voltage and low-power operation.  In particular, the optical performance is easily tuned via device geometry and doping concentrations of silicon and oxygen, making such devices promising candidates for high-speed and sensitive photodetection throughout the infrared where detectors are typically slow or require cooling.

\section*{Results and discussion}
In this section, we first show the fabrication process development to pattern TiSi$_x$O$_y$N$_z$ nanostructures. Then we discuss our design for electrically-connected bow-tie nanoantenna arrays exhibiting an enhanced optical response.  Finally, we report on the fabrication and testing of the optical and electronic characteristics of the bow-tie arrays, analysis of the data, and operation-induced damage.

\textbf{Fabrication process development.} Thin films of TiSi$_x$O$_y$N$_z$ were grown both on silicon and silicon oxide substrates by a reactive magnetron sputtering (MSP) co-deposition with the relative target powers tuned to modify dopant concentration. Details of this process can be found in Ref. \citep{19}. This procedure provides advantages over other growth methods for TiN optical dispersion engineering as it does not need modification of growth parameters that can induce unwanted film morphology nor does it require precise control of residual oxygen in the sputtering system's background vacuum \cite{Braic_titanium_oxynitride}. Likewise, the introduction of Si can be beneficial as Si$_3$N$_4$ phases have been shown to protect TiN phases from unwanted oxidation at high temperatures in oxygen-rich environments \cite{Veprek_superhard_composite}.
A 4:1 Ti target power to SiO$_2$ target power ratio was used to create thin films having a thickness of roughly \SI{50}{\nano\meter}.

We started our investigation on different etchants considering those already present in literature for the reactive ion etching (RIE) of TiN \cite{45,46,47,48,49}.  For dry etchants, we compared Cl$_2$, CF$_4$ and CF$_4$/O$_2$ plasmas. To evaluate the etch rate, we collected thickness measurements by ellipsometry after every minute of etching.

While Cl$_2$-based plasmas have been shown by others to achieve the highest etch rate for undoped TiN films, we found that Cl$_2$ plasmas did not effectively etch doped TiSi$_x$O$_y$N$_z$ films for any of the tested conditions (100 W, 10 sccm, 10 mTorr; 100 W, 10 sccm, 20 mTorr; 150 W, 20 sccm, 20 mTorr). The inability of chlorine to etch doped TiSi$_x$O$_y$N$_z$ films is likely due to the presence of Si-O and Ti-O bonds in the lattice of the film. Compound materials tend to etch atomically and as a consequence, we have found that the RIE of TiSi$_x$O$_y$N$_z$ is analogous to that of TiN with the addition of SiO$_2$ and TiO$_2$ passivation layers. While Cl$_2$ plasmas are a good RIE etchant for TiN, they are not as effective at etching SiO$_2$ \cite{46}. The etching procedure using Cl$_2$ might be improved by adding Ar ions into the gas flow to introduce a ballistic contribution to the etching process, but this could reduce the sidewall slope of the etched profile which is undesirable.

However, the TiSi$_x$O$_y$N$_z$ films were successfully etched by all of the tested fluorine-based plasmas (CF$_4$ and CF$_4$/O$_2$). With a pressure of 10 mTorr and a gas concentration of 15 sccm, CF$_4$ exhibited an etch rate of \SI{4.3}{\nano\metre\per\minute} when using a power of 100 W, and \SI{9.2}{\nano\metre\per\minute} at 200 W. By adding oxygen to the plasma, CF$_4$/O$_2$ achieved an etch rate of \SI{5.1}{\nano\metre\per\minute} with 10 mTorr, 9:1 sccm and 100 W, and an etch rate of \SI{9.4}{\nano\metre\per\minute} with a power of 200 W while maintaining the other parameters (for a visual comparison, see Figure S1 in the supplementary information). This slight increase in the etch rate was expected due to the presence of oxygen in the plasma, which increases the fluorine concentration in the chamber \cite{50}.

We also studied the sensitivity of the TiSi$_x$O$_y$N$_z$ films to typical wet etchants that might be used to remove hard masks after dry-etching for patterning. None of the tested etchants (buffered oxide etch (BOE) hydrofluoric acid (HF) 7:1, Transene aluminum etch type A, KMG CR7 chrome etch without surfactant, tetramethylammonium hydroxide (TMAH) and TMAH-based developers) effectively etched TiSi$_x$O$_y$N$_z$, with the only observed impact being a removal of the superficial oxide layer. This robustness provided flexibility in the choice of a suitable hard mask.

The process flow for the fabrication of a hard mask for patterning nanostructures into TiSi$_x$O$_y$N$_z$ is shown in Figure \ref{fig:Figure1}(a). We spin-coated about 80-nm-thick poly(methyl methacrylate) (PMMA) onto the sample and exposed the pattern by electron beam lithography (Extra High Tension (EHT): \SI{125}{\kilo\eV}) with a dose of about \SI{5000}{\micro\coulomb\per\centi\metre\squared} (see Figures \ref{fig:Figure1}(a.i-ii). Then we developed the resist and deposited by electron beam deposition a 30-\si{\nano\metre}-thick hard mask on the patterned resist (Figure \ref{fig:Figure1}(a.iii)). We performed the lift-off of the resist in heated N-Methyl-2-Pyrrolidone (NMP) in order to transfer a negative image of the resist. (Figure \ref{fig:Figure1}(a.iv)) We then etched the TiSi$_x$O$_y$N$_z$ film using dry RIE (Figure \ref{fig:Figure1}(a.v)). The etch time was calculated considering two minutes of over-etch with respect to the nominal etch rate evaluated for the film to ensure the gaps between the bow-ties are completely etched through.  We expect the over-etch to extend from 20 up to \SI{50}{\nano\metre} into the underlying substrate. To remove the hard mask, a suitable wet etchant was used to reveal the final device structure (Figure \ref{fig:Figure1}(a.vi)).

We evaluated the use of several different materials as hard masks: titanium, aluminum, molybdenum, and chromium. Chromium and aluminum achieved the best results in terms of gap dimensions (see Figure S2 in the supplementary information). In particular, chromium was found to be a very robust hard mask for CF$_4$/O$_2$ dry etching. The sidewall steepness of the gap was typically near \ang{90}. However, our standard chrome etch (KMG CR7) was not effective for dissolving the thin hard mask, presumably because of contamination induced in the hard mask by the etch process. While aluminum achieved good gap dimensions, the resulting TiSi$_x$O$_y$N$_z$ sidewall slope was not optimal. The taper angle in the case of aluminum was about \ang{70}.  However, unlike chromium, aluminum was easily removed using TMAH and TMAH-based developers. 

In the end, we achieved the best results by using a \SI{20}{\nano\metre} aluminum / \SI{10}{\nano\metre} chromium bilayer stack. The top layer of chromium acted as the hard mask during the RIE process, while the underlying aluminum layer was easily removed with a TMAH-based developer, enabling a clean lift-off of the hard mask after patterning. To test the performance of the bilayer hard mask, we first performed RIE using CF$_4$/O$_2$ 9:1 sccm, 100 W, 10 mTorr. The hard mask removal was then performed through a 15 min sonication in a TMAH bath. 

\begin{figure} [h!]
\centering
  \includegraphics[width=1\linewidth]{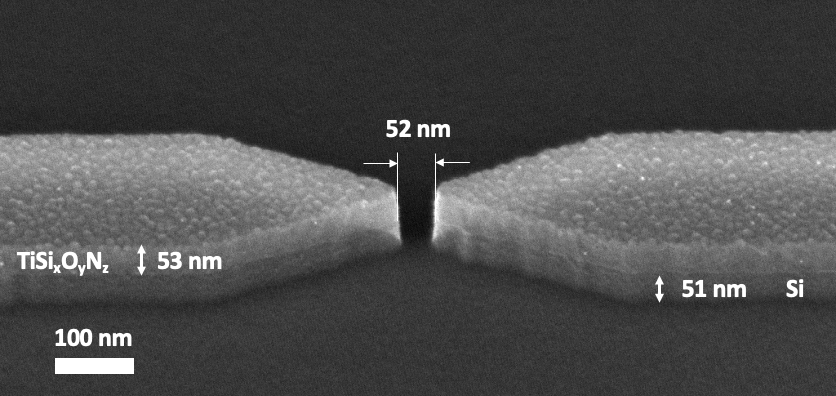}
  \caption{Scanning electron micrograph of fabricated TiSi$_x$O$_y$N$_z$ nanostructures on silicon after RIE and hard mask removal. The micrograph was taken with the sample tilted at \ang{45}.  The 51 nm overetched thickness in the silicon substrate is evident.}
  \label{fig:AlCr}
\end{figure}

Resulting nanostructures are shown in Figure \ref{fig:AlCr}. At the sidewalls, the contrast between the TiSi$_x$O$_y$N$_z$ layer and the underlying silicon substrate is evident. The etching time was calibrated with the purpose of overetching into the substrate, ensuring a completely etched profile inside the gap, and reducing the potential for charge accumulation in the substrate during electron emission testing. Gaps of about \SI{10}{\nano\metre} were obtained with this process, as shown in the nanoantenna reported in Figure \ref{fig:Figure1}(b). In that case, the ratio between the film thickness and the gap size was roughly $\sim$5. We also demonstrated that the same process can achieve sub-20-\si{\nano\metre} gaps even in the case of undoped titanium nitride (see Figure S3 in the supplementary information).

Reliable control over the gap dimensions was obtained both for TiSi$_x$O$_y$N$_z$ and undoped titanium nitride, which extends the use of the developed fabrication process to different areas of applications whenever sub-20-\si{\nano\metre} gaps are required. About 20 different samples were used to develop the fabrication process. The final process has been proved to be reliable thanks to the results obtained consecutively over about 5 samples on different substrates (\textit{i.e.} silicon and silicon dioxide), over a time of a few weeks.

\textbf{Design and fabrication of bow-tie nanoantennas.} A common nanostructure that enables large surface field enhancements in noble metals is the `bow-tie' structure. In this system, two triangularly-shaped nanostructures are oriented towards each other with a small gap separation that enables large surface field enhancements. Here, optical properties are highly dependent on nanostructure geometry. Specifically, gap-size is inversely correlated with field enhancement and therefore small gap size is desirable.  In this paper, we use this nanostructure type in order to form a direct comparison of fabricability, performance, and durability with existing noble metal field emission devices.

\begin{figure} [h!]
\centering
  \includegraphics[width=1\linewidth]{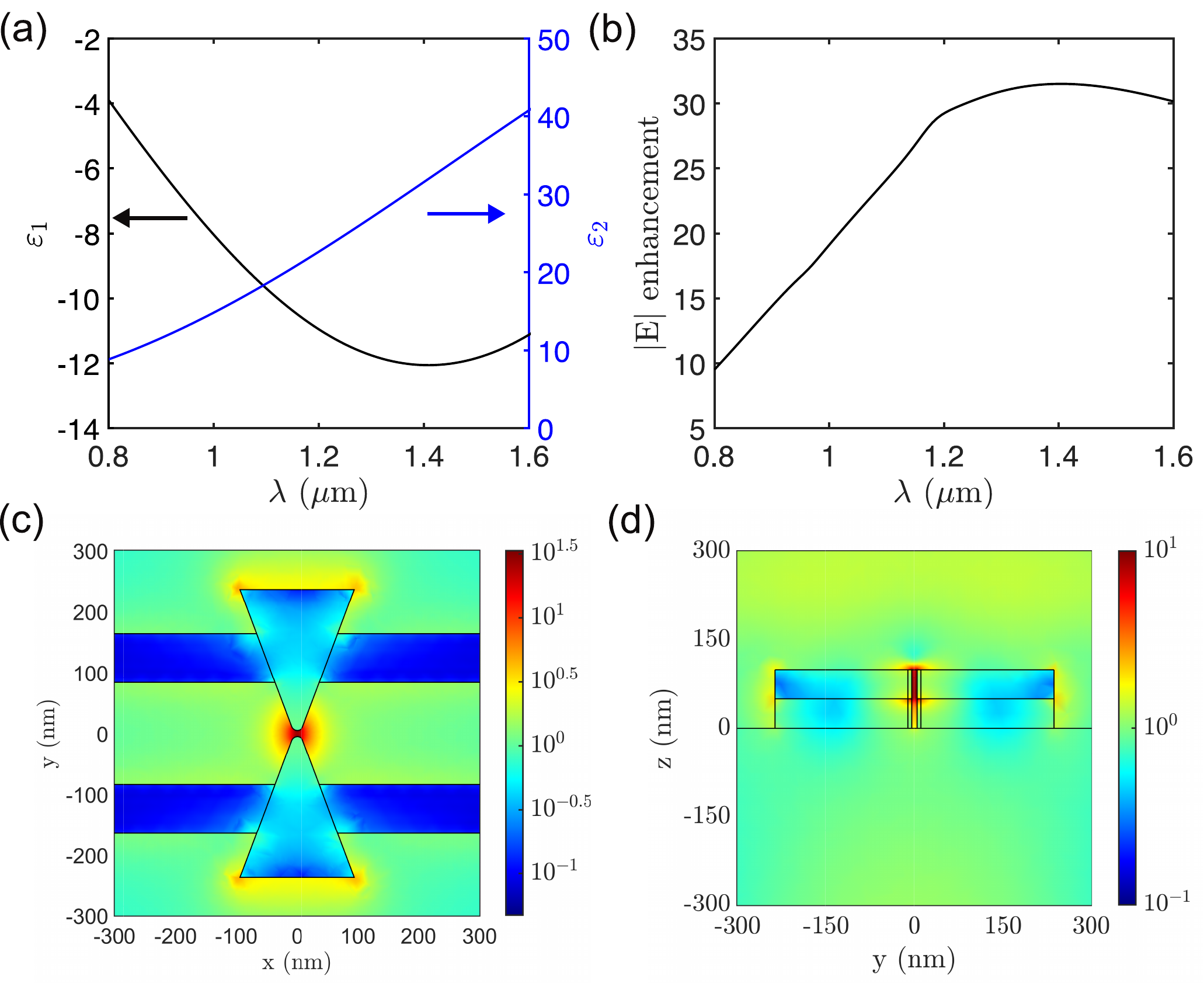}
  \caption{Optical properties of TiSi$_x$O$_y$N$_z$ thin films and nanostructures. (a) Real part $\epsilon_1$ and imaginary part $\epsilon_2$ of the material permittivity. (b) Field enhancement for a pair of bow-tie nanoantennas having the following parameters: triangle altitude of \SI{250}{\nano\metre}, triangle base width of \SI{188}{\nano\metre}, lead thickness of \SI{80}{\nano\metre}, TiSi$_x$O$_y$N$_z$ thickness of \SI{50}{\nano\metre}, and etched thickness into the oxide of \SI{50}{\nano\metre}). (c) The xy plane electric field profile at the middle of the bow-tie structure with $\lambda= \SI{1.6}{\micro\metre}$. (d) The xz plane electric field profile with $\lambda= \SI{1.6}{\micro\metre}$.}
  \label{fig:opt_properties}
\end{figure}

In order to accurately model the field enhancement of fabricated  TiSi$_x$O$_y$N$_z$ bow-tie nanoantennas, it was necessary to accurately characterize the dispersion of the deposited TiSi$_x$O$_y$N$_z$ thin films.  The optical dispersion of the TiSi$_x$O$_y$N$_z$ films used in this work was measured using variable angle spectroscopic ellipsometry and it is plotted in Figure \ref{fig:opt_properties}(a). We found that the deposition conditions resulted in a screened plasma wavelength of about \SI{635}{\nano\metre} and a transition to anomalous dispersion behavior at around \SI{1410}{\nano\metre}. Our ellipsometric modeling has excellent quality of fit that allows accurate characterization of film thickness even upon film etching.  

Using the dispersion data as shown in Figure \ref{fig:opt_properties}(a) we implemented numerical finite element method (FEM) simulations \cite{COMSOL} to evaluate the electric field enhancement that can be achieved in the gap of a bow-tie nanostructure system. We excite the bow-tie from the top surface with a plane wave. The simulated triangle height of each bow-tie element is \SI{250}{\nano\metre}, triangle base is \SI{188}{\nano\metre}, lead width is \SI{80}{\nano\metre} and the bow-tie thickness is \SI{50}{\nano\metre}. Considering the dry etching based process to pattern the nanostructures, a \SI{50}{\nano\metre} over-etch in the substrate has been taken into account and we used a curvature with \SI{10}{\nano\metre} radius at the corner of the bow-tie nanostructure in consideration of fabrication nonidealities. We show the electric field enhancement evaluated in the center of the gap in Figure \ref{fig:opt_properties}(b) with respect to the incident wavelength. Figures \ref{fig:opt_properties}(c) and (d) show the electric field profile when the incident wavelength is $\lambda$ = \SI{1.6}{\micro\metre} for a plane parallel to the $xy$ plane at the middle of the bow-tie nanostructure's thickness and at the $xz$ plane, respectively. Using these conditions, we achieve maximum field enhancements in the gap larger than 30. Although Au bow-tie nanostructures exhibit much higher simulated maximum field enhancement, TiSi$_x$O$_y$N$_z$ can counterbalance for performance values thanks to the realization of smaller, more stable, and more reproducible gap sizes. Likewise, the optimization by engineering TiSi$_x$O$_y$N$_z$ dispersion properties for field emission devices and investigation of structure types that better support LSP resonances are subjects of future study.

\begin{figure} [h!]
\centering
  \includegraphics[width=1\linewidth]{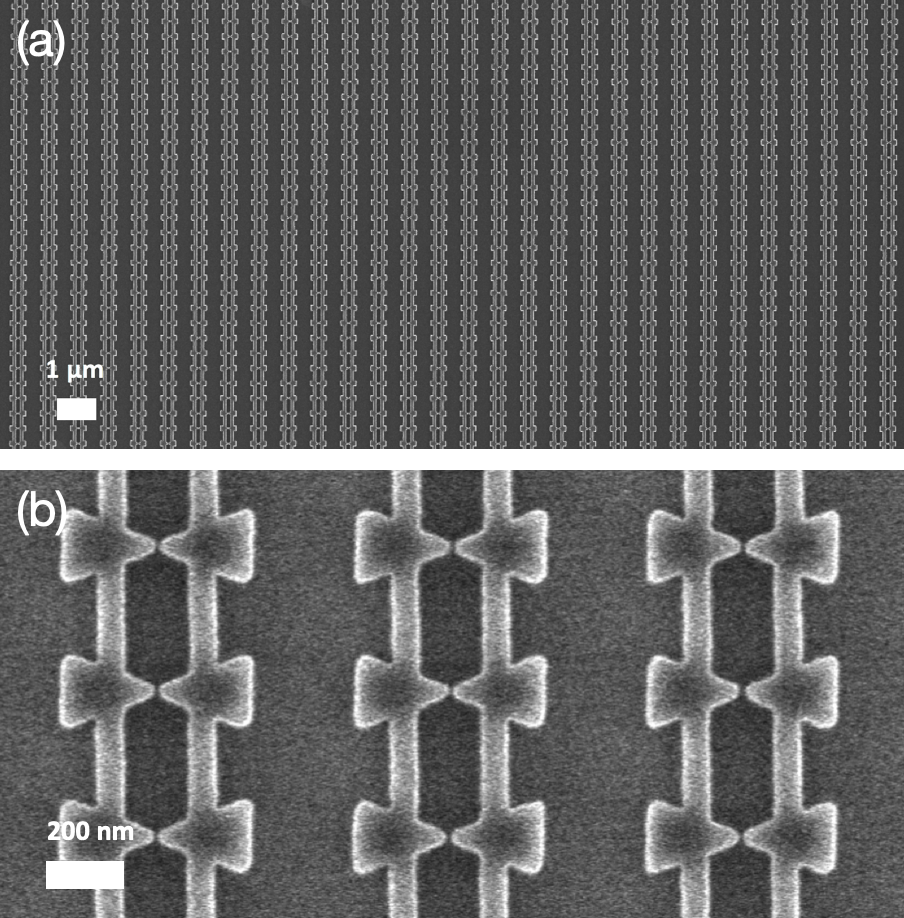}
  \caption{Top-down scanning electron micrographs of fabricated TiSi$_x$O$_y$N$_z$ bow-tie nanoantenna arrays. (a) Low magnification image of a nanoantenna array fabricated on silicon. (b) Close-up micrograph.}
  \label{fig:nanostrucures}
\end{figure}

By using the process presented in the previous section, we were able to pattern nanoantenna arrays with narrow gaps spanning several micrometers containing 2016 devices over an area of \SI{20}{\micro\metre} $\times$ \SI{20}{\micro\metre} (see \ref{fig:nanostrucures}(a)). The spacing between different columns is \SI{780}{\nano\metre}, whereas the vertical spacing between adjacent devices in the same column is \SI{400}{\nano\metre}. We patterned devices on SiO$_2$ and Si substrates. A close-up image of a fabricated array on silicon is shown in Figure \ref{fig:nanostrucures}(b). The devices patterned on SiO$_2$ were used for electronic and optoelectronic testing, while those patterned on Si were used for imaging purposes to better evaluate the etching process avoiding charging effects in the SEM. To ensure the same gap dimensions in the devices fabricated on the SiO$_2$ substrate as in the ones on Si, we included proximity effect corrections (PEC) during the EBL step.

\begin{figure} [h!]
\centering
  \includegraphics[width=1\linewidth]{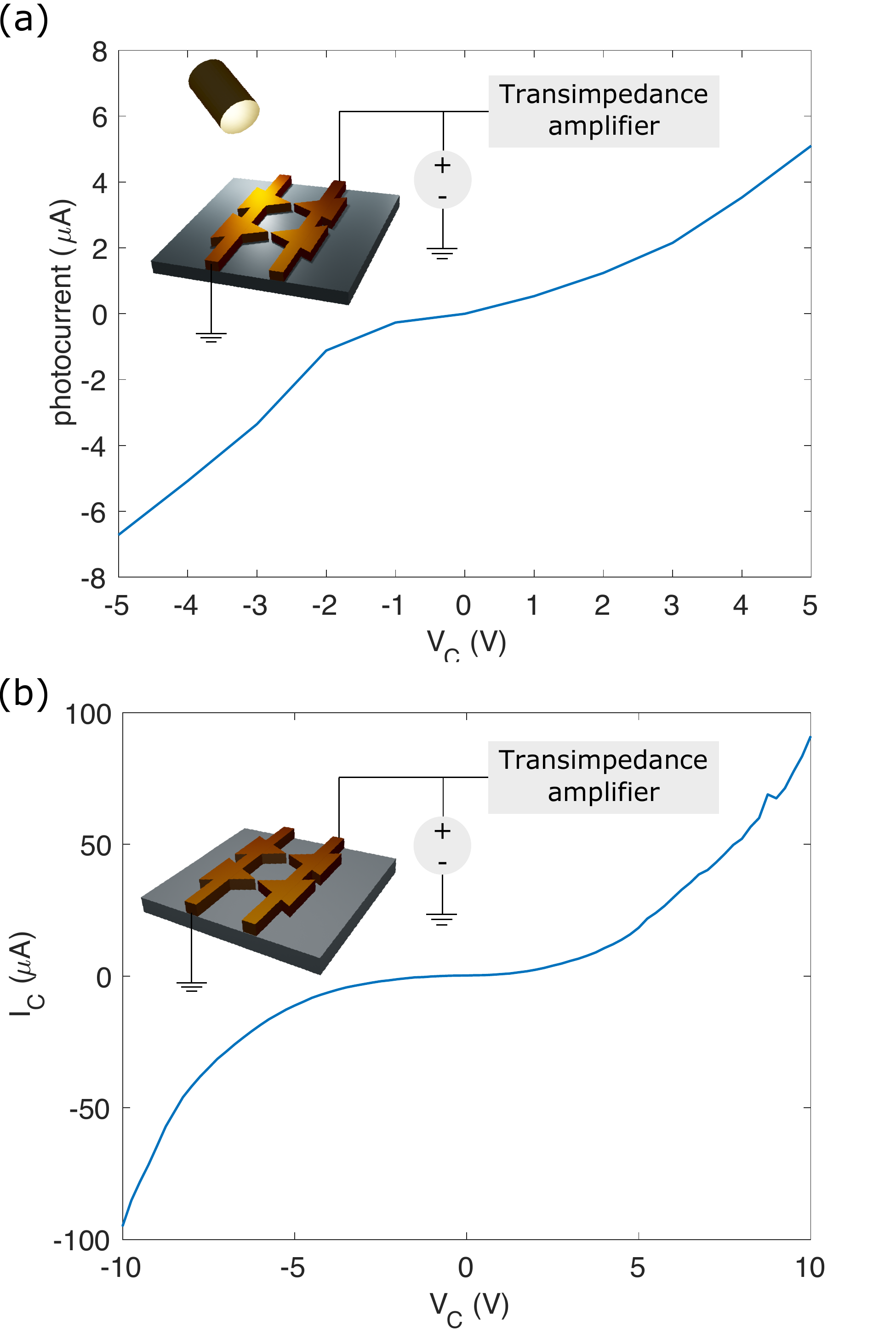}
  \caption{Optical and electrical characterization for a fabricated array composed of 2016 bow-tie nanoantennas with an average gap of \SI{14}{\nano\metre}. (a) Optical I-V curve with \SI{13}{\milli\watt} laser power optical excitation. The voltage sweep ranges from \SI{-5}{\volt} up to \SI{5}{\volt}. The elliptical laser spot was roughly \SI{2.4}{\micro\metre} $\times$ \SI{5}{\micro\metre} full-width at half maximum at the point of interaction with the device array. Inset: experimental setup composed of a supercontinuum broadband laser, a bias voltage source, a transimpedance amplifier. The lock-in amplifier is not shown in the picture. (b) Electrical I-V curve. The voltage sweep ranges from \SI{-10}{\volt} up to \SI{10}{\volt}. Inset: experimental setup composed of a bias voltage source, a transimpedance amplifier. The oscilloscope is not shown in the picture.}
  \label{fig:electrical_optical}
\end{figure}

\textbf{Optical characterization of devices.} To optically characterize the fabricated array, we used a setup like the one shown in the inset of Figure \ref{fig:electrical_optical}(a). To contact the devices, titanium/gold contact pads were fabricated using an additional photolithography step. The measured array (\SI{20}{\micro\metre} $\times$ \SI{20}{\micro\metre}) was composed of 2016 bow-tie nanoantennas, all simultaneously exposed to an external DC bias (Yokogawa GS200) to enhance the sensitivity of the nanoantennas and, thus, the emitted photocurrent \cite{turchetti2020impact}. We biased one side of the bow-tie array, while the other side was attached to ground. We illuminated the bow-tie array with broadband, supercontinuum laser source with 10-\si{\femto\second}-long pulses and a central wavelength of roughly \SI{1.2}{\micro\metre}, and a spectrum spanning $\sim$1-\SI{1.5}{\micro\metre} \cite{67}. To separate the photocurrent from the DC component due to the bias, we used a lock-in detection scheme. The current produced by the array is collected by a transimpedance amplifier (DDPCA-300). We fed the output of the transimpedance amplifier to a lock-in amplifier and modulated the laser amplitude with a chopper (at \SI{100}{\hertz}) placed in the laser path to isolate the photocurrent. 

Given the estimated dimension of the laser spot (\textit{i.e.} \SI{2.4}{\micro\metre}$\times$\SI{5}{\micro\metre}, each bow-tie nanoantenna of the array emitted an average current of \SI{150}{\nano\ampere}, when illuminated. The average quantum efficiency over the entire array was found to be around $0.5\times10^{-3}$ (\SI{0.5}{\milli\ampere\per\watt}), with a peak quantum efficiency of about $1.2\times10^{-3}$ (\SI{1.2}{\milli\ampere\per\watt}) at the most active spot in the array. This quantum efficiency is comparable to that of a bulk silicon photodiode at \SI{1.2}{\micro\metre}, but it is achieved from discrete nanoscale devices. TiSi$_x$O$_y$N$_z$ nanoantennas achieve a quantum efficiency that is roughly two orders of magnitude higher than the one obtained with Au nanostripes with vertical Al$_2$O$_3$ 4.5-nm-wide gap as shown in Ref. \citep{Chalabi_nanostripe}. This result is likely related to the long hot carrier lifetimes, which allow many more electrons to be collected by the drain. The TiSi$_x$O$_y$N$_z$ quantum efficiency could be further improved with further optimization of the optical design and/or material properties.  In Ref. \citep{8}, the same pulsed laser source and type of experimental setup were used to measure the response from an array of triangular Au emitters. With a bias voltage of \SI{30}{\volt}, a single nanotriangle produced an average current three orders of magnitude lower than the one we obtained from a TiSi$_x$O$_y$N$_z$ bow-tie nanoantenna.

\textbf{Electronic characterization of devices.} In this section, we discuss the electronic behavior of the fabricated device array when exposed to a DC bias. The experimental setup we used for electrical testing is shown in the inset of Figure \ref{fig:electrical_optical}(b). We biased one side of the bow-tie array, while the other side was attached to ground. We performed a voltage sweep in time and measured the output current using a transimpedance amplifier attached to an oscilloscope (LeCroy WaveSurfer 24Xs). 

We tested a nanoantenna array with average gap dimensions of \SI{14}{\nano\metre}. The electrical I-V characteristics are reported in Figure \ref{fig:electrical_optical}(b). The I-V curve is symmetric about the origin due to the symmetry of the bow-tie devices. Of particular interest is the low turn-on voltage (just a few volts on both sides), which is considerably lower than typical vertical emitters\cite{akinwande2000} and comparable with other planar emitters with similar geometries \cite{Nirantar2018}. The current levels are also comparable with these latter planar devices. In fact, averaging the electrical current over the number of nanoantennas in the array, we estimate that each nanoantenna is driving a current of \SI{5}{\nano\ampere}, on average. These characteristics, together with the hardness of TiSiON, could allow the use of these devices in high-speed nano-vacuum electronics in harsh environments \cite{Li_refractory_tin,Nirantar2018,02}. We did not observe any degradation in the I-V curve over time after roughly 6 hours of operation. Note that the testing was carried out in ambient conditions (\textit{i.e.} in the atmosphere at room temperature). 

\textbf{Study of the device degradation after testing.} The large electric fields that the antennas are exposed to during operation can lead to material migration, and the emitted electrons bombarding the nanoantenna surfaces can lead to an increase of the gap size and degradation of the tip shape if the material is not robust. For instance, it has been recently shown that gold nanoantenna arrays have larger gaps after testing, and this increase in gap size leads to a reduction of the emitted current and thus the device sensitivity over time. \cite{Yang_Nat_comm}

To highlight the robustness of TiSi$_x$O$_y$N$_z$ tested here, we imaged the array before and after testing. Before testing, the average gap size was estimated to be 14 $\pm$ 2.2 \si{\nano\metre}. (see Figure S4(a) in the supplementary information). In the first few minutes of testing, we observed a slight drift in the emitted current before stabilizing. After testing for a few hours, the array was re-imaged. An example post-testing scanning electron microscope image is shown in Figure S4(b). The average gap shrank down to 10.6 $\pm$ 2.7 \si{\nano\metre}, and at several sparse locations, a few defects like the one at the top right of Figure S4(b) were also found, but such failures were uncommon. The physical reason behind this process is not fully clear yet, but we can conjecture a melting process due to the high temperature reached during the illumination under high optical fields. The same behavior has been observed in Au nanoantennas on sapphire when illuminated by intense optical fields. \cite{Sivis_UV_plasmonic}

\section*{Conclusions} 

We developed a reliable fabrication process for patterning nanostructures made of titanium silicon oxynitride. Using this process, we were able to achieve gaps ranging from 10 to \SI{15}{\nano\meter} having aspect ratios of roughly $\sim$5 and sidewall slope of $\sim$\ang{90}. While we focused on demonstrating the process using titanium silicon oxynitride in this work, we have also found the same process to be effective for fabricating nanostructures with similar gap sizes in bare titanium nitride films (see Figure S3 in the supplementary information). This flexibility will be important to future work using such TiN-based refractory material platforms where adjustments to silicon and oxygen doping concentrations as well as device geometries will be used to optimize device performance. 

Using this process, we fabricated \SI{20}{\micro\metre} $\times$ \SI{20}{\micro\metre} arrays of electrically connected nanoscale structures with few-nanometer vacuum channels. Such devices are attractive for emerging areas such as nanoscale vacuum electronics~\cite{Nirantar2018, 02}, and resonantly-enhanced, ultrafast photodetection~\cite{cleo,15,8,Rybka_subcycle_control}. We performed electrical and optical characterizations of the fabricated arrays, highlighting the ability to achieve stable nonlinear field-emission with sub-10-V turn-on voltages in ambient conditions, and a bias-enhanced optoelectronic response with peak quantum efficiencies of around $1\times10^{-3}$ at \SI{1.2}{\micro\meter} -- close to the performance of a bulk silicon photodiode, but from discrete nanoantenna devices having a film thickness of only \SI{50}{\nano\meter}. This quantum efficiency is roughly two orders of magnitude higher than that achieved with DC biased Au nanostructures present in literature \cite{Chalabi_nanostripe}. Moreover, with a bias voltage of \SI{10}{\volt}, a single TiSi$_x$O$_y$N$_z$ bow-tie nanoantenna produced an average current that was three orders of magnitude higher than the one produced by a triangular Au emitter with the same pulsed laser source and a bias voltage of \SI{30}{\volt} \cite{8}. Last, we showed a comparison of the devices before and after testing, in order to analyze the damage due to the measurements. The damage was limited, with a reduction in the average gap size from roughly \SI{14}{\nano\metre} to \SI{10}{\nano\metre}.  

Exploiting the developed process together with material engineering, plasmonic resonances having higher quality factors could be achieved, resulting in devices that are robust to high temperature and degradation. Further investigations of the effects of the precise silicon and oxygen doping concentration on the work function and electronic properties of the material are needed; also, optimization of the nanoantenna geometry and doping concentrations for maximizing the optoelectronic response of the devices is warranted.

\section*{Methods}
\noindent\textbf{FEM Simulations.} Scattering boundary conditions were imposed on the four lateral sides of the model. Perfectly matched layers with \SI{500}{\nano\metre} thickness are implemented on the top and the bottom of the model. The minimum mesh size used in the simulation was \SI{4}{\nano\metre} and the total degrees of freedom of the FEM simulation is 4,800,000. 

\noindent\textbf{Sample cleaning.} The standard solvent cleaning for the sample consists of an acetone, methanol, and isopropanol cleaning routine, followed by an N$_2$ blow drying process. Prior to starting the fabrication process, the sample was also cleaned with an oxygen plasma ashing process at \SI{50}{\watt} for 60 seconds.

\noindent\textbf{Electron beam lithography process.} We used PMMA A2 resist. We span it at 2.5 krpm and baked it at \SI{180}{\celsius} for 2 minutes. The resist thickness was about \SI{80}{\nano\metre}. We performed electron beam lithography to pattern nanostructures on the resist. The system used is Elionix ELS-F-125 equipped with an EHT of \SI{125}{\eV}. The current used during the exposure was \SI{2}{\nano\ampere}. Proximity effect correction (PEC) was carried out (automatically by software) to take into account the substrate contribution on backscattered electrons, which introduce changes in the dose needed during the exposure in different areas of the sample. The resist was developed in a 3:1 IPA:MIBK solution at \SI{0}{\celsius} for 60 seconds. 

\noindent\textbf{Resist lift-off.} After the hard mask evaporation, the PMMA resist was lifted off in heated NMP at 60-\SI{70}{\celsius} for 1 hour. 

\noindent\textbf{Ti/Au pads fabrication.} We fabricated Ti/Au pads by direct writing laser (DWL) photolithography. A double-layer resist was used to achieve better resolution during the lift-off of the contact pads. We spun PMGI SF-9 at 4.5 krpm and baked at \SI{180}{\celsius} for 90 seconds. Then we spun S1813 at 4.5 krpm and baked at \SI{100}{\celsius} for 90 seconds. We exposed the double-layer resist using a DWL with \SI{10}{\watt} at 35\% transmission. We developed the resist in MF CD-26 for 80 seconds and soaked in DI water for 1 minute. We evaporated by PVD \SI{50}{\nano\metre} Ti / \SI{250}{\nano\metre} Au. We lifted off the double-layer resist by soaking the sample in acetone for 15 minutes and then sonicating in acetone for 1 minute. We removed the PMGI SF-9 residue by soaking the sample in MF CD-26 for 2 minutes before rinsing it in DI water for 1 minute.  Finally, the sample was dried using compressed N$_2$.

\begin{acknowledgement}
This work was supported by the Air Force Office of Scientific Research under award numbers FA9550-19-1-0065, and FA9550-18- 1-0436. LDN acknowledges the partial support of the NSF program “Tunable Si-compatible Nonlinear Materials for Active Metaphotonics” under Award No. DMR 1709704 and of the Army Research Laboratory (ARL) under Cooperative
Agreement Number W911NF-12-2-0023. We would also like to thank the internal reviewers Akshay Agarwal and Owen Medeiros of the Quantum Nanostructures and Nanofabrication group at MIT.
\end{acknowledgement}

\begin{suppinfo}

Additional etch calibration graph and scanning electron micrographs, including nanostructures patterned with aluminum hard mask, nanostructures patterned with chromium hard mask, undoped titanium nitride nanostructures, nanoantenna array before and after testing.

\end{suppinfo}

\bibliographystyle{acm}
\bibliography{achemso}

\end{document}


\setcounter{figure}{0}
\renewcommand{\thefigure}{S\arabic{figure}}

\begin{figure} [h!]
\centering
  \includegraphics[width=1\linewidth]{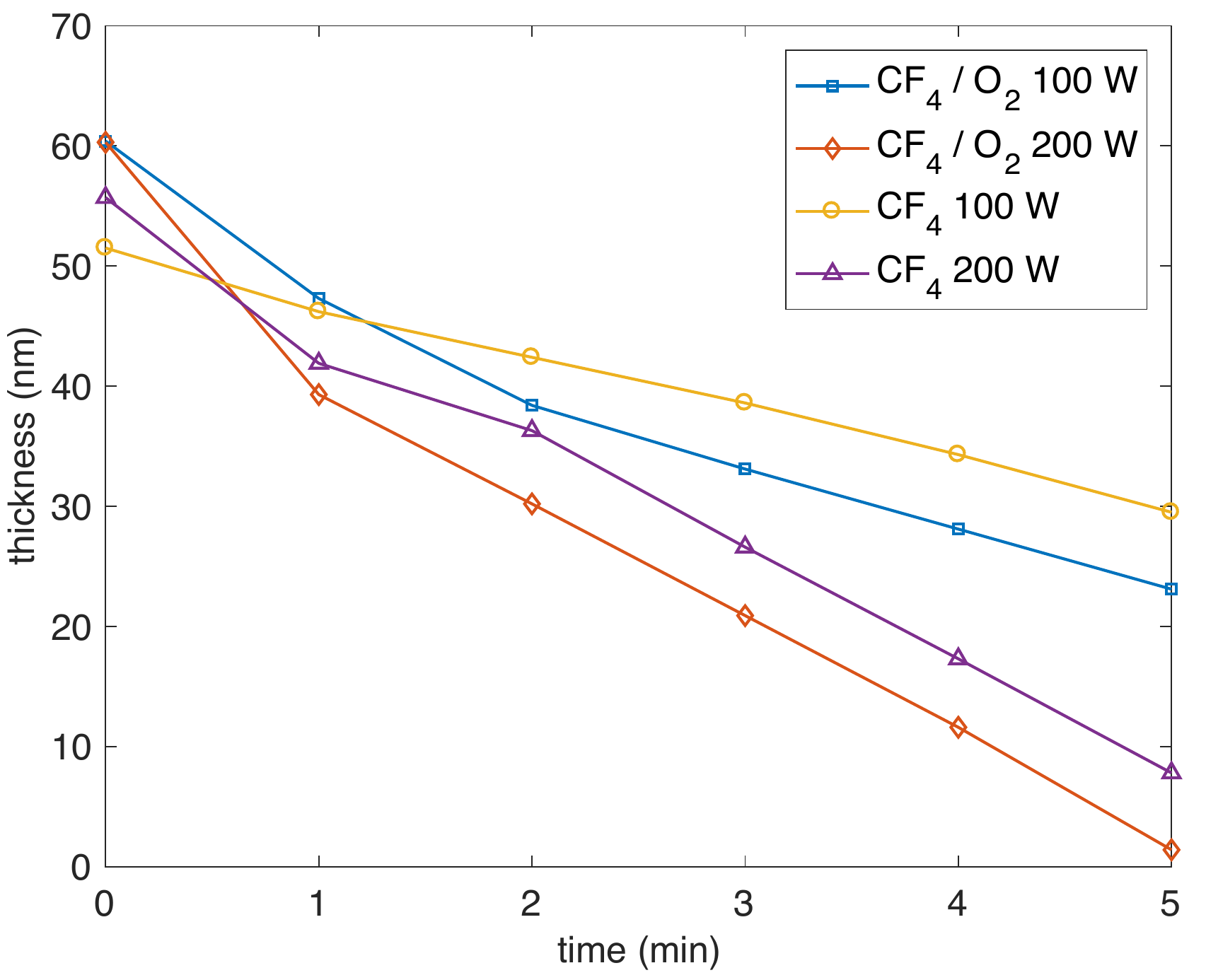}
  \caption{Etch rate comparison for different dry etching processes. CF$_4$/O$_2$ (9:1 sccm, 10 mTorr) and CF$_4$ (15 sccm, 10 mTorr) plasmas are shown with different powers: 100 W and 200 W. All the thickness measurements were collected by ellipsometry after every minute of etching. The starting thickness is not the same for all the processes because the samples came from different sputtering batches. A common trait is the larger drop in thickness within the first one to two minutes of each etch process.  We attribute this larger drop in thickness at the beginning of each etch process to the removal of the oxidized surface layer of the film.}
  \label{fig:etch_calibration}
\end{figure}

\begin{figure} [h!]
\centering
  \includegraphics[width=1\linewidth]{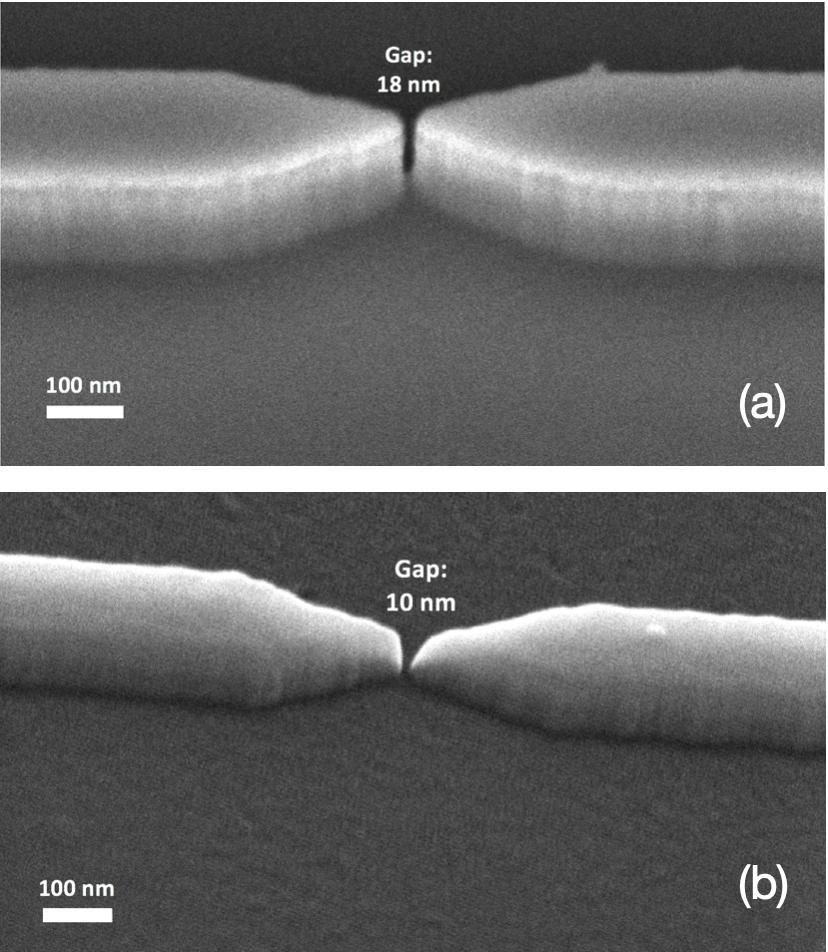}
  \caption{\ang{45}-tilted scanning electron micrographs of test structures for hard mask characterizations. (a) Chromium hard mask. Narrow gaps are achieved with very steep sidewalls. Chromium is still present on the structures even after Cr-7 etchant. The hard mask seems to have diffused into the underlying material, preventing the removal. (b) Aluminum hard mask. Narrow gaps are achieved, but the sidewalls have a taper angle of about \ang{70}, which makes the gap dimension estimate challenging. A thin gold layer is sputtered on the structures for imaging. Due to the thin layer of gold, the material stack is not visible.}
  \label{fig:chromium_aluminum}
\end{figure}

\begin{figure} [h!]
\centering
  \includegraphics[width=1\linewidth]{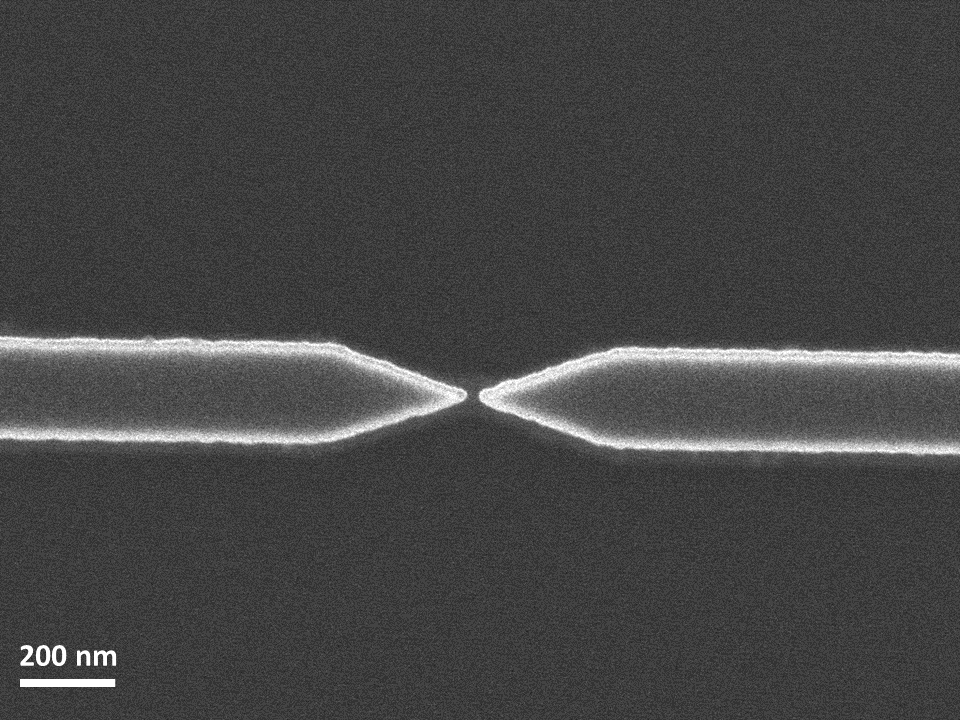}
  \caption{Top down scanning electron micrograph of an undoped TiN device.}
  \label{fig:tin}
\end{figure}

\begin{figure} [h!]
\centering
  \includegraphics[width=1\linewidth]{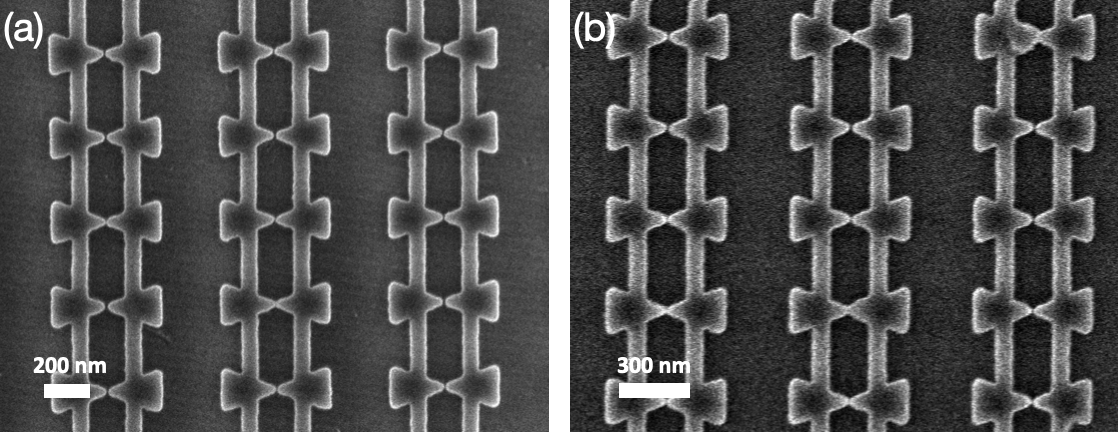}
  \caption{Top down scanning electron micrographs of the measured array. The two images show two different regions of the array. (a) Before testing. Average gap dimension: 14 $\pm$ 2.2 nm. (b) After testing. Average gap dimension: 10.6 $\pm$ 2.7 nm. A defect is present at the top right of the image, but such failures were not typical over the array.}
  \label{fig:before_after}
\end{figure}